\documentclass[10pt]{iopart}
\usepackage{epsfig}
\usepackage{supertabular}

\begin{document}

\title {Data Traffic Dynamics and Saturation on a Single Link}
\author {Reginald D. Smith}

\address{Bouchet-Franklin Research Institute, P.O. Box 14610
,Rochester, NY, 14610}
\date{February 19, 2009}
\ead {rsmith@sloan.mit.edu}


\begin{abstract}
The dynamics of User Datagram Protocol (UDP) traffic over Ethernet
between two computers are analyzed using nonlinear dynamics which
shows that there are two clear regimes in the data flow: free flow
and saturated. The two most important variables affecting this are
the packet size and packet flow rate. However, this transition is
due to a transcritical bifurcation rather than phase transition in
models such as in vehicle traffic or theorized large-scale computer
network congestion. It is hoped this model will help lay the
groundwork for further research on the dynamics of networks,
especially computer networks.
\end{abstract}

\maketitle

\section{Introduction}

The 1969 the Internet (then ARPANET) was first established as a
distributed packet communications network that would not only
reliably operate if some of its nodes were destroyed in an enemy
attack, but allow easier communications of computer research results
by universities. Today the Internet has grown to become a sprawling
network of every aspect of humanity dwarfing previous technological
mediums in both complexity and behavior. It was therefore only a
matter of time that advanced statistical techniques, such as those
developed by physicists in statistical mechanics were applied to
investigate it.

Since the late 1990s, the Internet has been of Interest to the
physics community, becoming aware to most in the seminal Nature
paper of Watts and Strogatz \cite{watts1}. This was continued or
paralleled by the work of countless others \cite{barabasi1,
newman1,newman2, dorogovtsev1,faloutsos, vespignani1} . However,
until recently this research has focused mostly on the topological
aspects of networks and much less on dynamics. A particularly
fertile area on network dynamics, and one related to this paper is
the study of phase transitions from free flow to congestion in
computer networks
\cite{internetphase1,internetphase2,internetphase3,internetphase4}.
Most results give a critical packet flow on networks which separates
free flow from congested traffic. There have been some
investigations of dynamics aspects such as synchronization of
coupled oscillators \cite{nonosc1,nonosc2,nonosc3,nonosc4} and some
metabolic dynamics \cite{metdyn1,metdyn2} as well thus dynamics is
rapidly moving from being a peripheral to a primary discussion about
networks.

A very interesting reverse situation is visible in the studies of
the statistical mechanics of vehicular traffic. Vehicle traffic on
roads has been investigated, also using statistical mechanics, but
focusing on the dynamics and flow of traffic versus the topology of
the road network \cite{traffic1,traffic2,traffic3}. Though there is
some overlap between these two topics, traffic flow has been
described in many ways with the most common description using the
fundamental diagram of vehicle flow vs. vehicle density. Most models
propose a two or three phase model of traffic. The first phase is
free flow, where cars drive near the speed limit with little
congestion and influence on each other's velocity. The final phase
is congested traffic where traffic flow becomes spontaneously
congested after reaching a critical density. In the three phase
model, there is an intermediate phase called synchronized flow where
traffic is not congested but cars match their speed at a reduced
speed level effectively increasing the correlation length of the
system as a prelude to congestion. In \cite{trafficcompare}
Ga\'{a}bor and Csabai conduct a similar study comparing vehicle
traffic flow to data packet flow using the number of TCP connections
as the variable for flow density. They find that a fundamental
diagram like pattern appears in data traffic when the flow density
is modeled as the number of TCP connections between two different
endpoints.

\section{Internet Traffic Research Issues}

In investigating Internet dynamics, one is struck by how similar the
dynamics of the network can be superficially similar to vehicle
traffic. Internet networks are made of the flow of countless packets
across network links and can be prone to the same free flow or
congestion that vehicle traffic can be. Like traffic data, data on
Internet traffic is easily obtainable either by setting up packet
sniffers and traffic analyzers between certain nodes or using public
data sets such as the WIDE Project's MAWI data traces from
trans-Pacific US-Japan T1 lines \cite{mawi} or the Stanford Linear
Accelerator Center (SLAC)'s PingER project which has monitored ICMP
ping response against different nodes across the Internet on a
continuous basis for years \cite{pinger}.

However, Internet traffic data analysis is complicated by many
factors that are not easily accountable for in theoretical models.
First, since the Internet is a decentralized network based on
dynamic routing, the route of the traffic is not completely
transparent. The path from one point to another can be fairly fluid
and changing even within a continuous flow of transmitted data. Many
of the intermediate routers or autonomous systems (AS), which are
basically Internet service providers, have private and constantly
changing routing rules and configurations that alter traffic in
unpredictable ways to obtain certain quality of service metrics or
traffic shaping priorities \cite{aspaper}. This makes analysis of
data collected from Internet traffic fraught with questions and
confusion as far as disambiguating the effect of network topology on
dynamics.

In addition to topological constraints, the dynamics of Internet
traffic itself can affect dynamics studies. Under the Internet
Protocol (IP) suite, there are many transport level protocols such
as TCP and UDP and countless application level protocols such as
HTTP among others. Protocols at all levels, including transport and
application levels, can influence data traffic. For example, TCP has
a congestion control algorithm which will actually throttle network
speed given the feedback it receives from packet loss data on the
network, requires periodic acknowledgements from the destination
before sending more data, and will buffer data to send depending on
the round trip time (RTT) of the connection in order to guarantee
delivery \cite{tcp1, tcp2}. Therefore, measurements of TCP/IP
network speeds, even in relatively "clean" networks, can actually be
extremely complicated and dependent on much more than the topology
of the network or volume of traffic.

Finally, the statistical nature of Internet traffic is still poorly
understood. Internet traffic volumes and packet interarrival times
to not follow typical distributions in other communications networks
such as Poisson or Erlang distributions but rather exhibit bursty,
self-similar traffic patterns which have been very difficult to
model and predict \cite{multifrac1, multifrac2, multifrac3,
multifrac4}. Like in measurements of Internet topology where a
relatively small amount of nodes have many edges, almost everything
in the Internet that can be measured seems to have a long-tailed
distribution as a matter of course. TCP and UDP traffic flows follow
this trend where a relatively small number of flows carry to bulk of
data transferred (so-called "elephant flows")
\cite{mawielephant,mawielephant2}.The origins of these patterns of
traffic are still a matter of research and debate. Add to this other
inherent uncertainties and patterns in Internet traffic flow such as
trimodal distributions of packet sizes \cite{packetsize}, traffic
spikes to due to malicious code such as viruses or Trojan directed
distributed denial-of-service attacks \cite{caida1,caida2}, and
periodicities in the volume of traffic caused by 12 hour, 24 hour,
and 7 day cycles (with a 3.5 day harmonic)
\cite{periodicity1,periodicity2}. There is also an issue of
long-range correlations of Internet traffic between different
routers which can be relatively uncorrelated or very correlated
depending on the nature of traffic and the level of congestion
\cite{trafficcorrelations}. All of these factors are mentioned to
demonstrate that modeling and understanding Internet traffic
dynamics is a problem likely of greater magnitude than topological
analysis. Given these difficulties and more, understanding the
basics of traffic dynamics and the interactions between topology and
dynamics in computer networks is essential to understanding theoretical aspects, creating accurate simulations, and conducting useful experiments.

\section{Experimental Setup}

The purpose of the experiment described by this paper is to ask a
basic question about traffic dynamics: how is the throughput (speed)
of a link affected by three fundamental variables that determine the
nature of network traffic: the average packet size, the average
packet flow rate, and the bandwidth of the link. The bandwidth of
the network, in this case 100 Mbps (megabits per second), is the
theoretically maximum possible throughput. The throughput itself can
be described in terms of the average packet size and average flow
rate by
\begin{equation}
\label{eq1}
 \langle{T}\rangle = \langle{p}\rangle\langle{\lambda}\rangle
\end{equation}
Where $\langle{T}\rangle$ is the average throughput of the link, p
is the average size of packets in a transmission over a given period
of time and $\langle{\lambda}\rangle$ is the average flow rate in
packets per second. In this experiment since all packets will be off
the same size in each sample we can say
\begin{equation}
\label{eq2}
\langle{T}\rangle = p\langle{\lambda}\rangle
\end{equation}
Two computers, both running Windows XP, were connected using an Cat
5e cable that connected to the Ethernet network adapter (NIC) of
both computers. The Ethernet flow control was disabled to ensure
that the flow of data, and not signaling between the computers to
prevent dropped packets that flow control entails, determines the
throughput. Traffic was generated using the program Iperf which
generates a stream of identically sized packets at a throughput
inputted by the user. The throughput in Iperf was designated at 100
Mbps in order to test what the maximum throughput the network would
actually demonstrate given an attempt at maximum throughput. The
transport protocol UDP was chosen over TCP for several reasons. UDP
is a connectionless protocol, meaning it does not guarantee
delivery, and will solely submit a string of packets. TCP in a
connection based protocol whose delivery guarantee requires frequent
"handshaking" between the source and destination and whose
congestion control algorithm can affect throughput in a non-trivial
fashion delivering a lower throughput due to protocol software, not
the maximum throughput of the link. Finally, TCP can give different
performance over links with different latencies, as measured by
packet RTT, so the results may not be subject to larger
generalization \cite{tcp1}.

\begin{figure}

    \centering
    \includegraphics[height=0.5in, width=1.5in]{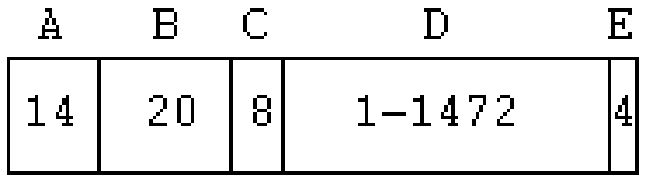}

        \caption{Structure of a packet in this paper. Proportions based on a 50 byte payload. Numbers are size of headers or
        payload in bytes. A is the Ethernet header which contains MAC address source and destination and payload type,
        B is the Internet Protocol (IP) header, C is the User Datagram Protocol (UDP) header, D is the
        data payload and E is the Ethernet CRC checksum hash to prevent accidental corruption of the frame.}
    \label{overhead}
\end{figure}

To test the performance of the network under different sized
packets, the UDP packet payload was varied from 25 bytes up to 1450
bytes in 25 byte increments. In order to ensure that equation
\ref{eq2} holds, you must ensure that all packets are the same size.
The structure of packets under Ethernet/IP is shown in figure
\ref{overhead}. The payload, a variable selected in the Iperf
software, is encapsulated by a header for UDP (8 bytes), IP (20
bytes), and a header and footer in the Ethernet frame (total 18
bytes). Frame is just a general term for an Ethernet packet.
These headers mostly provide routing data, priorities, checksums,
and other information important to packet logistics. In addition, in
standard Ethernet the maximum frame size, minus Ethernet headers and
footers, is 1500 bytes. With the Ethernet overhead the total maximum
size for an Ethernet frame is 1518 bytes. Therefore, at 1475 bytes
payload, the total would be 1503 bytes and you would have packet
fragmentation - instead of one frame you would have two, one with a
frame payload of 1500 bytes and a second with a frame payload of 3
bytes. This would affect throughput by showing a sudden change since
average frame payload size would drop to about 750. Given the
phenomenon of fragmentation, an actual analysis of the effect of
packet size on throughput is only useful up the fragmentation size
limit.

In the experiment, Iperf delivered a packed stream of UDP traffic
from the client to the server computer, trying to send as close to
bandwidth as possible, and outputted the average throughput in Kbps
(kilobits per second). It also gave the packet loss, and a measure
called jitter which is not used but measures the deviation in packet
interarrival times versus interdeparture times. For throughput,
Iperf measures a related measured called goodput which measures data
speed in terms of the payload size, not including any packet
overhead in the calculation of bytes transferred. However, the
packet flow rate is accurate and is calculated from equation
\ref{eq2}. Therefore the actual throughput was recalculated using
packet sizes that include both payload and packet overhead.

\section{Results}
\begin{figure}

    \centering
    \includegraphics[height=2in, width=2in]{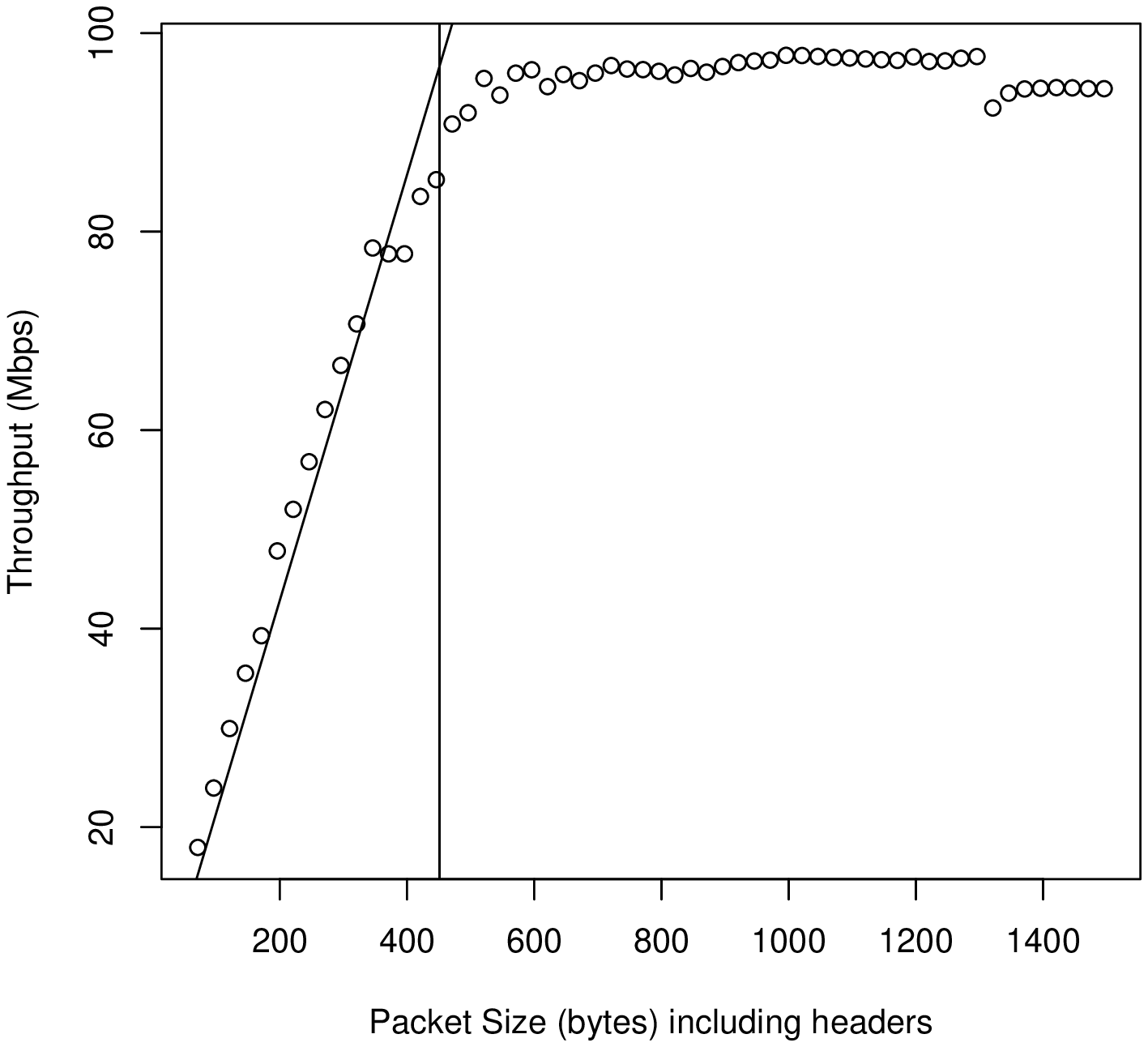}
    \includegraphics[height=2in, width=2in]{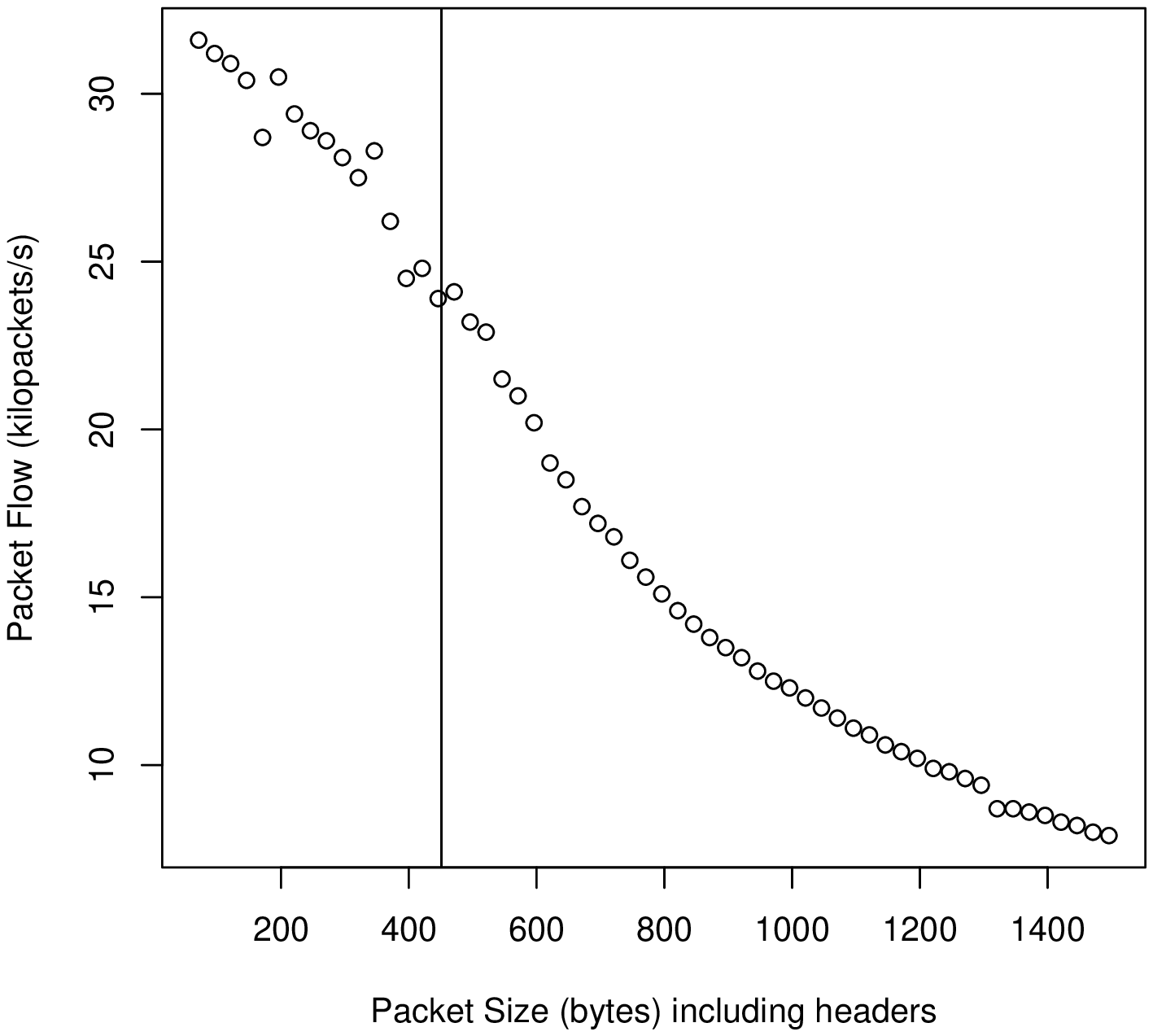}
    \includegraphics[height=2in, width=2in]{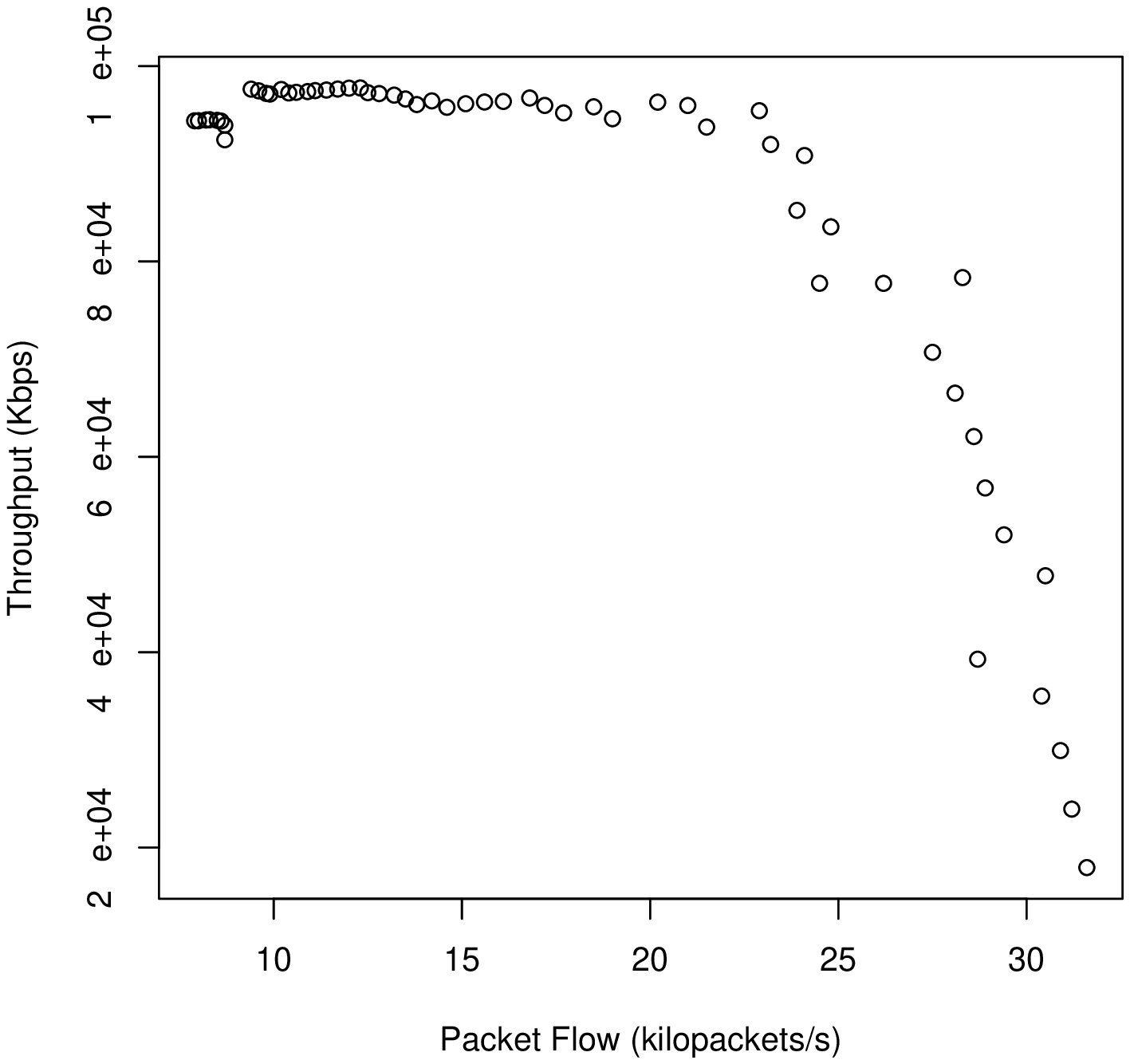}
        \caption{Graphs of the average throughput vs. packet size, average packet flow vs. packet size, and average throughput vs. average packet flow respectfully. Vertical lines
        on the first two graphs represent the calculated critical packet size of 451 bytes.
        The fitted line on the first graph is the line predicted by maximum packet flow. The slight decrease at high packet
        size in the first graph is due to unknown system effects and not inconsistent enough to reject experimental results.}
    \label{graphs1}
\end{figure}
At all packet sizes, packet loss was small, much less than 1\%. The
results of the experiment are shown in figure \ref{graphs1}. First,
it is clear that throughput decreases with decreasing packet size.
This first fact is well-known in the network engineering community
\cite{MTU1}. This is an inherent property of all network adapters,
Ethernet or otherwise. In fact, one of the key requirements in next
generation networks is the network capability to send "jumbo frames"
where fragmentation limits at the network layer are much larger than
1500 bytes, up to 9000 bytes in some cases. These larger packet
sizes cause higher throughput and more efficient networking because
within the network adapter and computer hardware, there is a
per-packet processing overhead.

Despite the bandwidth rating of network adapters, be it 10,100, or
1000 Mbps, there is a maximum packet flow that they can effectively
handle. Given equations \ref{eq1} and \ref{eq2} it is clear that to
maintain any given throughput, by lowering the packet size you are
increasing the packet flow. Because of the packet flow processing
bottleneck in the hardware, however, this can make high throughput
impossible at low packet sizes.

As seen in figure \ref{graphs1}, for large packet sizes, the
throughput is very close to bandwidth and can be roughly equivalent
to free flow traffic. At a specific critical packet size, however,
$p_{c}$, the throughput begins to rapidly degrade to the point it is
only about 6\% of bandwidth at 25 bytes. This is the saturated
state. Here saturation is used instead of congestion since
congestion is usually a network wide phenomenon while this packet
slowdown in throughput is due to overwhelming the processing power
at the NIC. In a superficial way this behavior is similar to the
fundamental diagram flow-density curve in vehicle traffic. Packet
flow increases with smaller packet size until saturation forces
packet flow to begin to slow its increase and approach a maximum
value. Comparisons between vehicle traffic should be qualified
though. In vehicle traffic, there is interaction between cars on the
road giving rise to the collective dynamics which justify a
statistical mechanics interpretation. On data networks, packets do
not interact with each other and packet collisions are errors, not
intrinsic aspects of the packet flow. Therefore, as will be shown
below, the transition from free flow to saturation should be viewed
as a bifurcation in the system dynamics, not as a phase change. On
the network level where there are many interacting nodes, perhaps
congestion can be seen as a phase change but this perspective is not
appropriate at the single link level.

As stated earlier, in free flow the throughput is nearly bandwidth
and comparatively, though not completely, steady state. In this
region, there is a mutual relationship between packet size and
packet flow. Differentiating equation \ref{eq2} we have

\begin{equation}
\label{tradeoff} d\langle{T}\rangle = pd\langle{\lambda}\rangle +
\lambda dp
\end{equation}

assuming $d\langle{T}\rangle$ is 0 in free flow regardless of the
packet size or flow we can conclude

\begin{equation}
\label{tradeoff2}
\frac{dp}{d\langle{\lambda}\rangle}=-\frac{p}{\langle{\lambda}\rangle}
\end{equation}

So there is a tradeoff curve, like the production possibility
frontiers in economics, between packet size and flow in free flow
traffic. Also,

\begin{equation}
\label{tradeoff3}
\frac{dp}{P}=-\frac{d\langle{\lambda}\rangle}{\langle{\lambda}\rangle}
\end{equation}

demonstrating that every increase in packet flow is matched by a
corresponding decrease in packet size and vice versa. Because the
networking and computer equipment have various processes and
imperfections our free flow region never reaches bandwidth and
steadily erodes with smaller packet size, however, the high
throughput feature is relatively constant compared to the saturated
state.

At a packet size $p_{c}$, we have an increasingly rapid breakdown in
throughput. This corresponds approximately to the maximum flow to
the network adapter and the breakdown into saturation. We calculate
the theoretical maximum flow as

\begin{equation}
\label{maxflow}
\langle{\lambda}\rangle_{c}=\frac{\langle{T}\rangle_{max}}{p_{c}}
\end{equation}

where in the theoretically ideal situation $\langle{T}\rangle_{max}
= B$ where $B$ is the bandwidth. Therefore in the saturated region,
the throughput $\langle{T}\rangle$is given by

\begin{equation}
\label{congestedflow}
\langle{T}\rangle=\frac{p}{p_{c}}\langle{T}\rangle_{max}
\end{equation}

so the ratio of packet size to the critical packet size determines
the throughput under saturation compared to the maximum possible
throughput. In the first graph of figure \ref{graphs1}, is a
comparison of this prediction with the observed data in the
saturated region where $\langle{T}\rangle_{max}$ is about 96 Mbps.
Given data from the saturated region for $\langle{T}\rangle$,
$\langle{T}\rangle_{max}$, and $p$ we can estimate the critical
packet size $ p_{c}$. In this case, the critical packet size is
approximately 451 bytes at around 25k packets/s. Though this model
seems to accurately predict the throughput values for almost the
entire region of saturation, there is a seemingly glaring
contradiction in the second graph of figure \ref{graphs1} where the
packet flow keeps increasing with decreasing packet size. Though the
packet flow does not stagnate, as it should according to ideal
theory, the third graph shows that the decline in throughput over a
fairly short range of packet flow demonstrates packet flow is the
limiting factor in the saturated region. In the congested region the
packet flow increases from about 25 to 30 thousand flows per second
which is a definite increase but relatively small compared to the
region of free flow when it increased from 7 to about 25 without
affecting throughput. Note, $p_{c}$ is specific to the equipment and configuration used and does not have a general value of 451 bytes.

\section{Bifurcation Analysis}
\begin{figure}

    \centering
    \includegraphics[height=2in, width=2in]{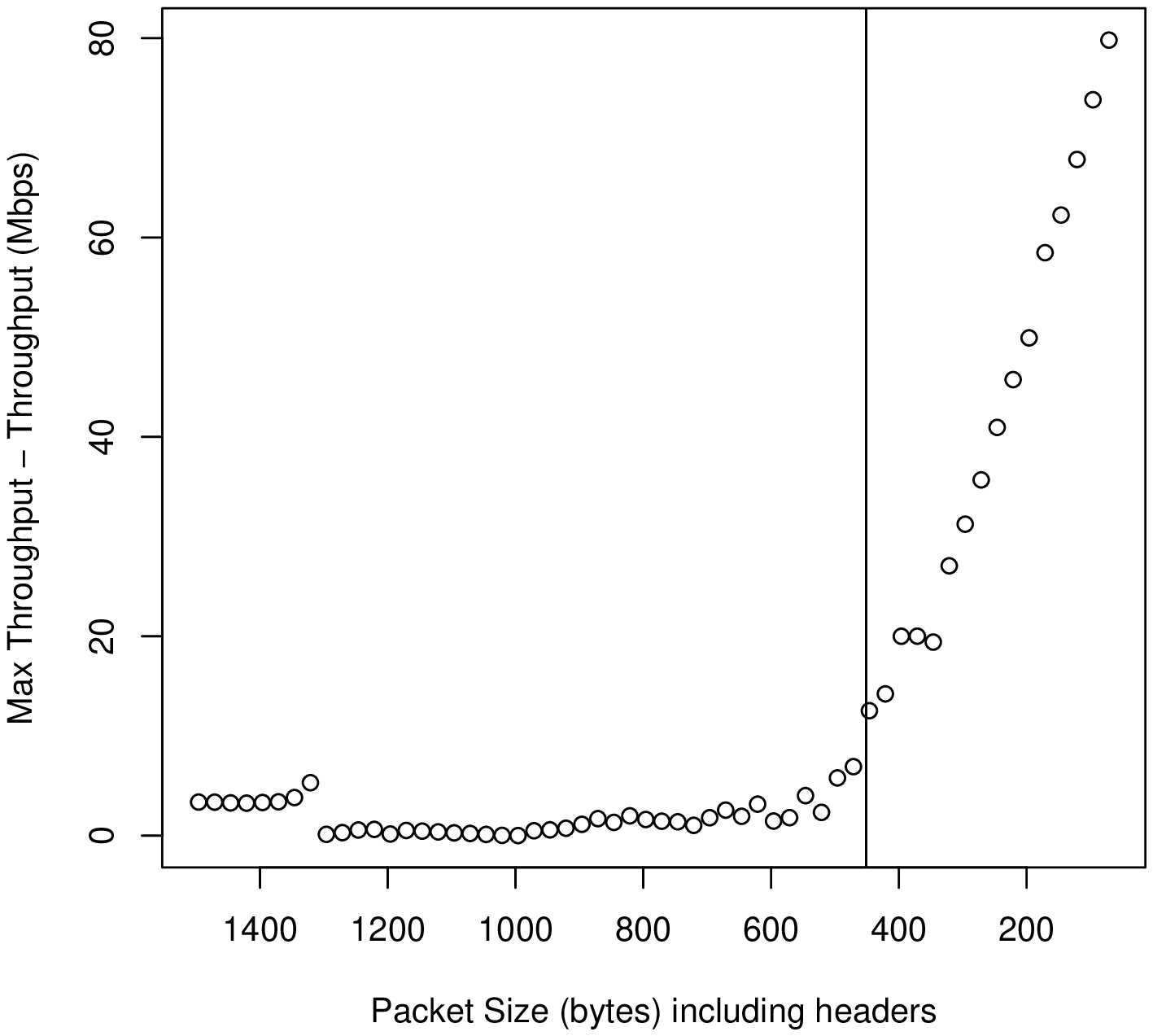}

        \caption{Graph of the maximum throughput measured minus throughput vs packet size on a reverse axis. The estimated critical packet
        size is given by the vertical line.}
    \label{bifurcationgraph}
\end{figure}

The fact that the stable throughput of the system changes with the
parameter of packet size leads us to suspect a possible bifurcation.
The exchange of stability between quasi-bandwidth throughput and
packet flow limited throughput leads to the hypothesis of a
transcritical bifurcation with stability changing at $p_{c}$.
Assuming $\langle{T}\rangle_{max} = B$ the two stable throughput
regimes are $\langle{T}\rangle = B$ and $\langle{T}\rangle =
(p/p_{c})B$, so

\begin{equation}
\label{bifurcation} d\langle{T}\rangle/dt =
(\langle{T}\rangle-B)(\frac{p}{p_{c}}B - \langle{T}\rangle)
\end{equation}

and

\begin{equation}
\label{bifurcation2} d\langle{T}\rangle/dt = -\frac{p}{p_{c}}B^{2}
+\langle{T}\rangle{B}(1+ \frac{p}{p_{c}}) - \langle{T}\rangle^{2}
\end{equation}

This can be seen as the traditional form of a transcritical
bifurcation where $dx/dt = px-x^2$, however, this is made more clear
by making the independent variable $B - \langle{T}\rangle$:

\begin{equation}
\label{bifurcation3} d(B-\langle{T}\rangle)/dt =
B(1-\frac{p}{p_{c}})(B-\langle{T}\rangle) -(B-\langle{T}\rangle)^2
\end{equation}

which fits the normal form for a transcritical bifurcation, $dx/dt =
px-x^2$. The transcritical bifurcation can be more clearly seen in
the graph of $(\langle{T}\rangle_{max} - \langle{T}\rangle)$ vs. the
reverse order axis with $p$ in figure \ref{bifurcationgraph}.

\section{Discussion}

As mentioned before, the transition from free flow to saturated
traffic here is a bifurcation not a phase change given the lack of
interaction among the constituent particles in the system. This
conclusion can also give pause to extrapolations of "classical" flow
on network theory to complex networks such as computer networks. If
looking at the weights and capacities of the network from the
perspective of throughput, typical maximum flow algorithms such as
max-flow min-cut may give incorrect answers if the limiting aspect
of the flow is the packet flow rate, size of the packets, or another
issue that is not plainly visible. Flows in computer networks do not
behave like incompressible fluid or similar flows assumed in most
flow models where any flow rate freely flows up to the capacity
minus costs incurred by friction, etc.

Another point is other similarities to vehicle traffic dynamics.
Interestingly, the first graph of Figure \ref{graphs1} is similar to
a flow-density curve seen in traffic models. Even equation \ref{eq1}
has similarity to such traffic models where
\begin{equation}
\label{trafficequation}
Q = DV
\end{equation}

Where $Q$ is the traffic flow in cars/h and $D$ is traffic density
in cars/km and $V$ is flow velocity in km/h. There is a critical
traffic density that separates free flow from synchronized flow or
congestion that plays a similar role to packet size in data
networks.

This paper does not address the larger problem of dynamics on
networks, however, it is the contention of this paper that
understanding the simple dynamics at the network level is essential
to understanding the wider implications of network sized dynamics.
Given the problems with using Internet traffic data described
earlier, more understanding in the
basics of traffic dynamic may be obtained through experimental
setups or computer network simulations. Future research should look
at how topological invariants in networks affect dynamics and how
dynamics may affect the evolution of networks and changes in their
topological invariants.

As a final note regarding research on dynamics in networks,
particularly data networks and the Internet, the author believes
that more interaction and cross-referencing between the engineering
and physics communities will help promote better understanding and
advancement. Although there is research at an almost feverish pitch
in both the physics and engineering community on networking, both
sets of publications seem to be almost entirely ignorant of each
other. In physics, one of the only such publications commonly cited
is the Faloutsos team's work on the router topology of the Internet
\cite{faloutsos}. Engineering researchers, in journals such as those
published by IEEE and ACM, also have rarely quoted physics
literature besides the most well-known papers of Barab\'{a}si,
Watts, or Strogatz and seem largely unaware of the later results
being reached by physics in the topology of networks. Without a
detailed understanding of the network protocols and engineering
literature regarding the function of data networks and the Internet,
this paper would not have been possible. A similar situation is seen
in sociology where a wealth of research has been done on aspects of
social network dynamics such as the diffusion of trends along
networks, in a quantitatively rigorous fashion, but again both
communities do not usually inform each other of progress or motivate
collaboration. As the study of networks expands and matures, it will
become necessary to read and reach across interdisciplinary
boundaries to share tools and knowledge that will allow the most
profound and predictive insights to be reached.

\end{document}